\documentclass[aps,prd,twocolumn,amsmath,amssymb,superscriptaddress, showkeys, showpacs, 10pt]{revtex4}

\usepackage{amssymb,amsmath,graphicx,color,microtype}

\usepackage{setspace}

\def \der{{\rm d}}

\def \nt{n_{\rm t}}

\begin{document}

\title{Precision of future experiments measuring primordial tensor fluctuation}


\author{Yi Wang}
\email{yw366@cam.ac.uk} \affiliation{Centre for Theoretical
Cosmology, DAMTP, University of Cambridge, Cambridge CB3 0WA, UK}

\author{Yin-Zhe Ma}
\email{mayinzhe@phas.ubc.ca} \affiliation{Department of Physics
and Astronomy, University of British Columbia, Vancouver, BC, V6T
1Z1, Canada;}

\begin{abstract}

Recently the second phase of Background Imaging of Cosmic
Extragalactic Polarization (BICEP2) claimed a detection of the
tensor-to-scalar ratio ($r$) of primordial fluctuation at
$5\sigma$ confidence level. If it is true, this large and
measurable amplitude ($r \simeq 0.2$) of B-mode polarization
indicates that it is possible to measure the shape of CMB B-mode
polarization with future experiments. 
We forecast the precision of $r$ and the tensor spectral index 
$\nt$ measurements, with $\nt$ as a free parameter, 
from a {\it Planck}-like experiment, and from Spider and 
POLARBEAR given the current understanding of their experimental noise and foreground contamination.
We quantitatively
determine the signal-to-noise of the measurement in $r$-$\nt$
parameter space for the three experiments. The forecasted
signal-to-noise ratio of the B-mode polarization somewhat depends
on $\nt$, but strongly depends on the true value of $r$.

\end{abstract}


\maketitle


{\it Introduction--} Recently the BICEP2 experiment claimed a more
than $5\sigma$ detection of CMB B-mode polarization \cite{BICEP2}.
This detection, if confirmed by ongoing and forthcoming
experiments, implies a large amplitude of primordial tensor
fluctuations and therefore has profound theoretical implications.
For instance, given the current detected amplitude $r=0.2$, the
inflationary potential and the associated derivatives can be
completely reconstructed around a few number of e-folds
\cite{Ma:2014vua}. However, on the other hand, several other
groups claimed recently that the BICEP2 results may come from the
spurious signal of the polarized dust~\cite{Flauger14}.

Assuming the BICEP2 result is correct and therefore the primordial
tensor fluctuation is measurable, it is possible to measure not
only the amplitude but also the shape of the primordial tensor
power spectrum with future experiment. The BICEP2 measured B-mode
power spectrum has power excess at small scales, indicating a blue
tilt ($\nt$) of the spectrum \footnote{The preliminary
cross-correlation between BICEP2 and the Keck array does not
support such power excess. However, the BICEP2-Keck cross
correlation has power deficit at large scales, which also slightly
prefers a blue tensor spectrum; For other possible explanations, see, for example, refs.~\cite{Xia:2014tda, Cai:2014xxa, Cai:2014bea}.}.~
The statistical significance of
such a blue tensor spectrum is found to be in between $1\sigma$
and $2\sigma$ \cite{Gerbino:2014eqa, Wang:2014kqa}.

The hint of blue $\nt$ becomes stronger when the BICEP2 data is
combined with {\it Wilkinson Microwave Anisotropy Probe}
(\textit{WMAP}) and \textit{Planck} data~\cite{Wang:2014kqa,
Ashoorioon:2014nta, Smith:2014kka}. In fact, before the tensor
mode is detected, the theoretical prediction of temperature power
spectrum is around $5\%$--$10\%$ higher than the measurement on
$\ell<50$ \cite{Ade:2013kta}. The detected tensor-to-scalar ratio
$r=0.2$ will further enhance the low-$\ell$ temperature power
spectrum ($C^{\rm TT}_{\ell}$) by $10\%$ since the primordial
gravitational wave preserves only on very large scales. This
ensures that the standard model even more inconsistent with the
observational data.

The possibility of a blue power spectrum with positive $\nt$ can
reconcile the tension between model and the data. With positive
$\nt$, the contribution to $C^{\rm TT}_{\ell}$ ($\ell<50$) is less
than red tensor spectrum, making the model prediction more
consistent with the data~\cite{Liu04}. It has been shown that once $\nt$ is
released to be a free parameter in the likelihood analysis, a
positive $\nt$ is found to be at $3\sigma$ confidence level (CL)
\cite{Smith:2014kka, Cai:2014hja}. A similar hint for a blue power spectrum is
also found in the results of global fittings \cite{Wu:2014qxa,
Li:2014cka}.

Given the current BICEP2 constraints on $r$ and $\nt$, in this
paper we will investigate how precisely the on-going and future
experiments can measure these two parameters and therefore
determine the tensor spectrum. 
Specifically, we will forecast the precision of measurement from a {\it Planck}-like 
full-sky CMB experiment~\footnote{We use the {\it Planck} noise model in~\cite{Planckblue} 
to forecast the constraints, which may differ from the real {\it Planck} experiment noise model. So here we call it a {\it Planck}-like experiment, though for brevity it will continue to be labelled `{\it Planck}' in most of the following discussion.}, and from the Spider~\cite{Crill08} 
and POLARBEAR~\cite{PolarBear} experiments with the current understanding of their experimental noise and foreground contamination.

The primordial tensor power spectrum can be expanded in power law form:
\begin{eqnarray}
P_{\rm t}(k)=A_{\rm t}(k_{0})\left(\frac{k}{k_0} \right)^{\nt},
\label{eq:Pt}
\end{eqnarray}
where $k_{0}$ is the pivot wave number at which $\nt$ and $A_{\rm
t}$ are evaluated. The amplitude of tensor power spectrum, $A_{\rm
t}(k_{0})$ is related to the tensor-to-scalar ratio as given by
\begin{eqnarray}
r=\frac{A_{\rm t}(k_{0})}{A_{\rm s}(k_{0})},
\end{eqnarray}
where $A_{\rm s}(k_{0})$ is the scalar amplitude at $k_{0}$. In
our data analysis, we use $k_{0}=0.01 \,{\rm Mpc}^{-1}$. Then the
power spectrum $C_{\ell}^{\rm BB}$ is related to $P_{\rm t}(k)$
by
\begin{equation}
C_{\ell}^{\rm BB}=\frac{\pi }{4}\int P_{\rm t}(k)\Delta
_{\ell}^{\rm B}(k)^{2} {\rm d}\ln k, \label{clbb-eq}
\end{equation}%
where $\Delta^{\rm B}_{\ell}(k)$ is the transfer function for each
multipole $\ell$ which can be obtained from public code {\sc
camb}~\cite{cambcode}.

\begin{figure}[htbp]
  \centering
  \includegraphics[width=0.45\textwidth]{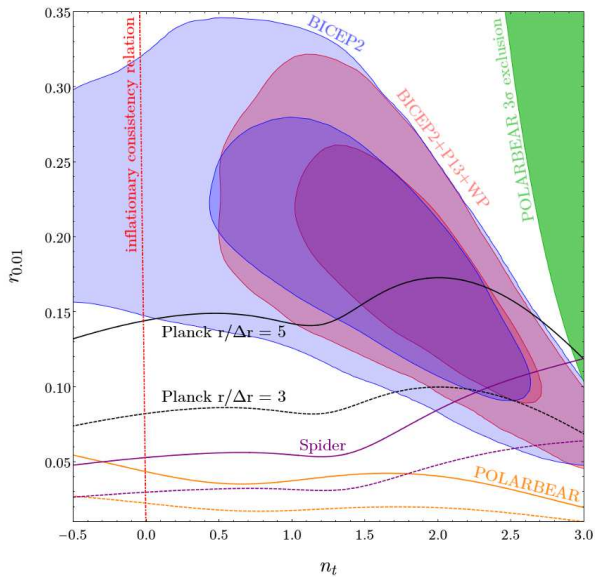}
  \caption{Signal-to-noise ratio of \textit{Planck} (black lines), Spider (purple lines) and POLARBEAR (orange lines) as functions of $r$ and $\nt$. Solid lines denote signal-to-noise ratio $r/\Delta r=5$ and the dashed lines denote signal-to-noise ratio $r/\Delta r=3$. Current observations are also plotted. Blue contours are the $1\sigma$ and $2\sigma$ constraints from BICEP2. Red contours are the combined constraints from BICEP2+\textit{Planck} (2013)+{\it WMAP} polarization (WP). Current POLARBEAR $3\sigma$ exclusion region is plotted in green. Nearly vertical red dash-dotted line is the consistency relation $r=-8\nt$ predicted by the minimal model of inflation.}
\label{fig:comb}
\end{figure}

{\it Constraining $r$ and $\nt$ from BICEP2 and \textit{Planck}
data--} Here we make use of the public code {\sc
CosmoMC}~\cite{Lewis:2002ah} to constraint $r$ and $\nt$. The
other cosmological parameters are fixed at the best-fitting value
from \textit{Planck}. With BICEP2 data and marginalizing over $r$,
we can obtain the likelihood on $\nt$ as $1.24\pm 0.90$
($1\sigma\,$ CL). By combining BICEP2 data with \textit{Planck}
(2013) and \textit{WMAP} polarization (WP) data, we find
$\nt=1.76\pm0.54$ ($1\sigma\,$\ CL). When marginalizing over
$\nt$, the likelihood on $r$ are $r=0.20\pm0.06$ and
$r=0.18\pm0.05$ at $1\sigma$ CL for BICEP2 only and
BICEP2+\textit{Planck} (2013)+WP, respectively. The $1\sigma$ and
$2\sigma$ contours are plotted on Figure~\ref{fig:comb}, where the
blue contours are for BICEP2 only, and the red contours are for
BICEP2+\textit{Planck} (2013)+WP. It is worth noting that the
inclusion of \textit{Planck} (2013)+WP does not change the contour
significantly at large positive $\nt$ part, but it sets strong
limit on small $\nt$. Therefore, the negative $\nt$ (red tilted
power spectrum) is disfavored at more than $3\sigma\,$CL. This is
clearly inconsistent with the consistency relation of the
single-field slow-roll inflation where $\nt$ is slightly negative
given the current measurement of $r$. If the positive $\nt$ is
found to be true, this clearly indicates some new physics for
inflation.

\begin{figure}
\centerline{
\includegraphics[width=3.6in]{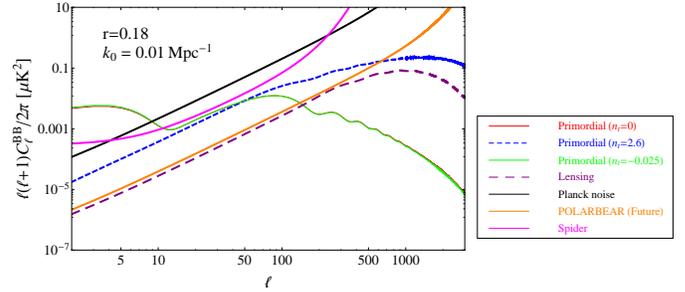}}
\caption{Comparison between B-mode polarization power spectra of
models with $\nt=0$ (flat power spectrum), $\nt=-r/8=-0.025$
(consistency relation) and $\nt=2.6$ ($2\sigma$ upper limit of
current constraint) and noise level for \textit{Planck}, Spider
and POLARBEAR experiments. Purple dashed line is the B-mode signal
induced by gravitational lensing (non-primordial).}
\label{fig:clbb1}
\end{figure}

\begin{figure}
\centerline{
\includegraphics[width=3.6in]{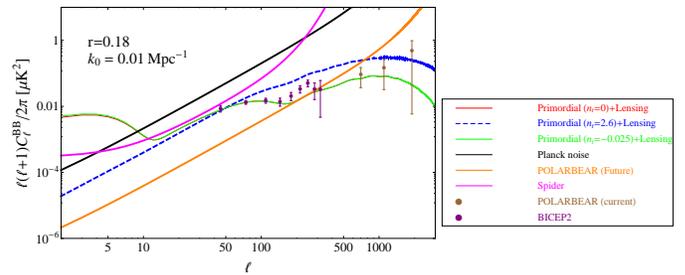}}
\caption{Comparison between the total B-mode polarization power
spectra (primordial plus lensing) of the same three models as
described previously and noise level for \textit{Planck}, Spider
and POLARBEAR experiments. Brown and purple data with error-bars
are the band-power data from the current experiments.}
\label{fig:clbb2}
\end{figure}

\begin{figure}
\centerline{
\includegraphics[width=3.8in]{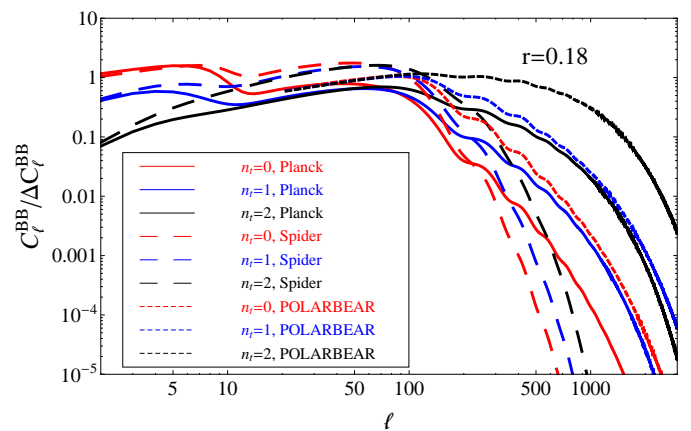}}
\caption{Signal to noise ($\Delta C_{\ell}^{\rm BB}$ is calculated
through eq.~(\ref{eq:delta-cl})) at each $\ell$ for
\textit{Planck} (solid lines), Spider (long-dashed lines) and
POLARBEAR (short-dashed lines). Value of $n_{\rm t}$ is taken to
be $0$ (red), $1$ (blue) and $2$ (black). Value of $r$ is fixed to
be $0.18$ at pivot scale $k_{0}=0.01\, {\rm Mpc}^{-1}$.}
\label{fig:snrcl}
\end{figure}

\begin{table}[tbp]
\begin{centering}
\begin{tabular}{|l|l|l|l|}\hline
Model & Experiment & $\ell \leq 100$ & $100< \ell < 3000$ \\
\hline & \textit{Planck} & 7.57 & 1.98 \\ \cline{2-4} $n_{\rm
t}=0$ & Spider & 14.89 & 4.18 \\ \cline{2-4} & POLARBEAR & 8.41 &
6.59
\\ \hline &
\textit{Planck} & 5.65 & 2.72 \\ \cline{2-4} $n_{\rm t}=1$ &
Spider & 12.87 & 5.28 \\ \cline{2-4} & POLARBEAR & 8.27 & 9.64
\\ \hline &
\textit{Planck} & 5.87 & 5.70 \\ \cline{2-4} $n_{\rm t}=2$ &
Spider & 13.48 & 8.55 \\ \cline{2-4} & PPOLARBEAR & 8.50 & 22.8
\\\hline
\end{tabular}%
\caption{Signal-to-noise ratio of the two bands for three
experiments, i.e., SNR=$\sqrt{\sum^{\ell_{\rm max}}_{\ell_{\rm
min}} (C^{\rm BB}_{\ell}/\Delta C^{\rm BB}_{\ell})^{2}}$}
\label{tab1}
\end{centering}
\end{table}

\begin{figure*}
\centerline{
\includegraphics[bb=0 0 476 321, width=3.2in]{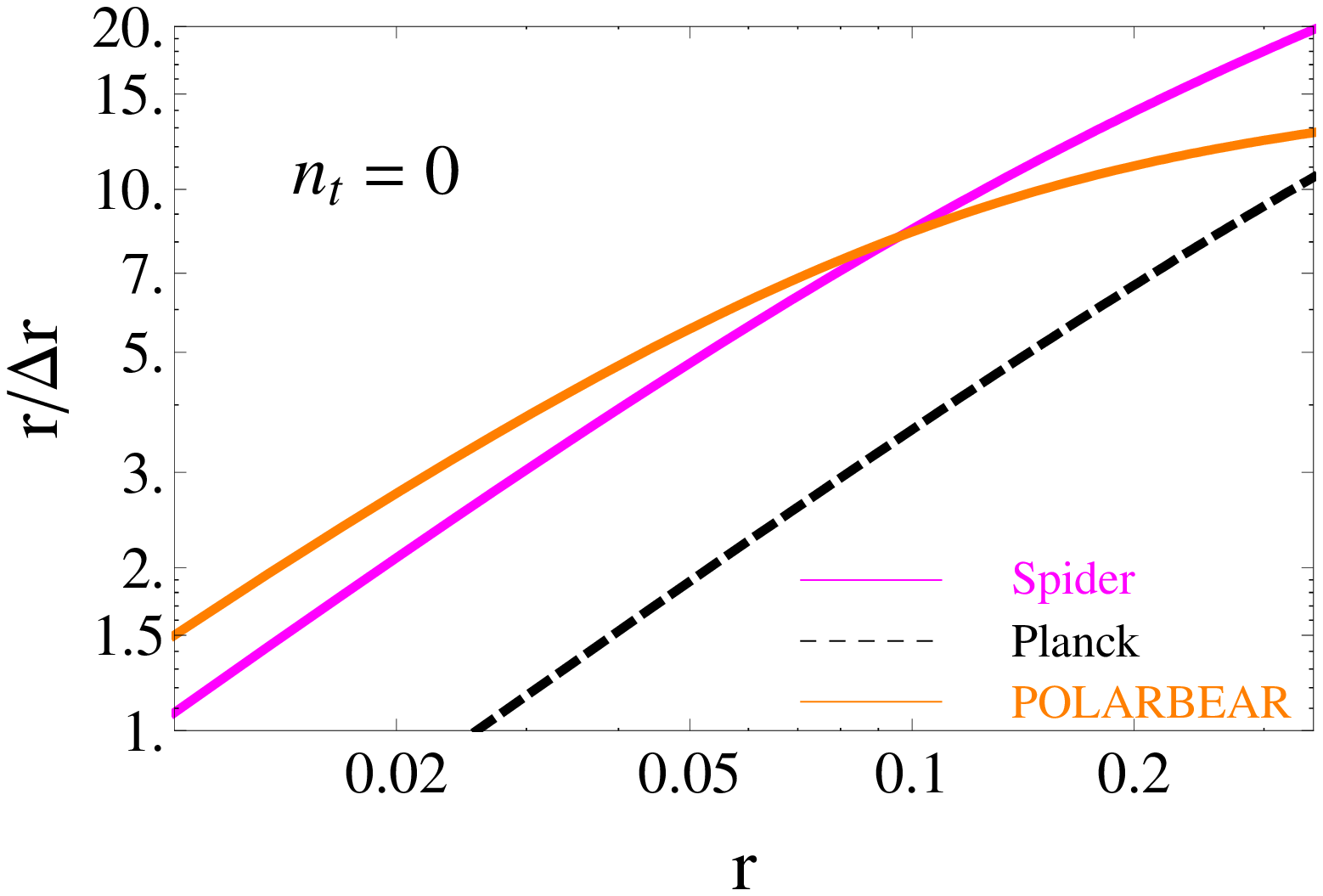}
\includegraphics[bb=0 0 480 323, width=3.2in]{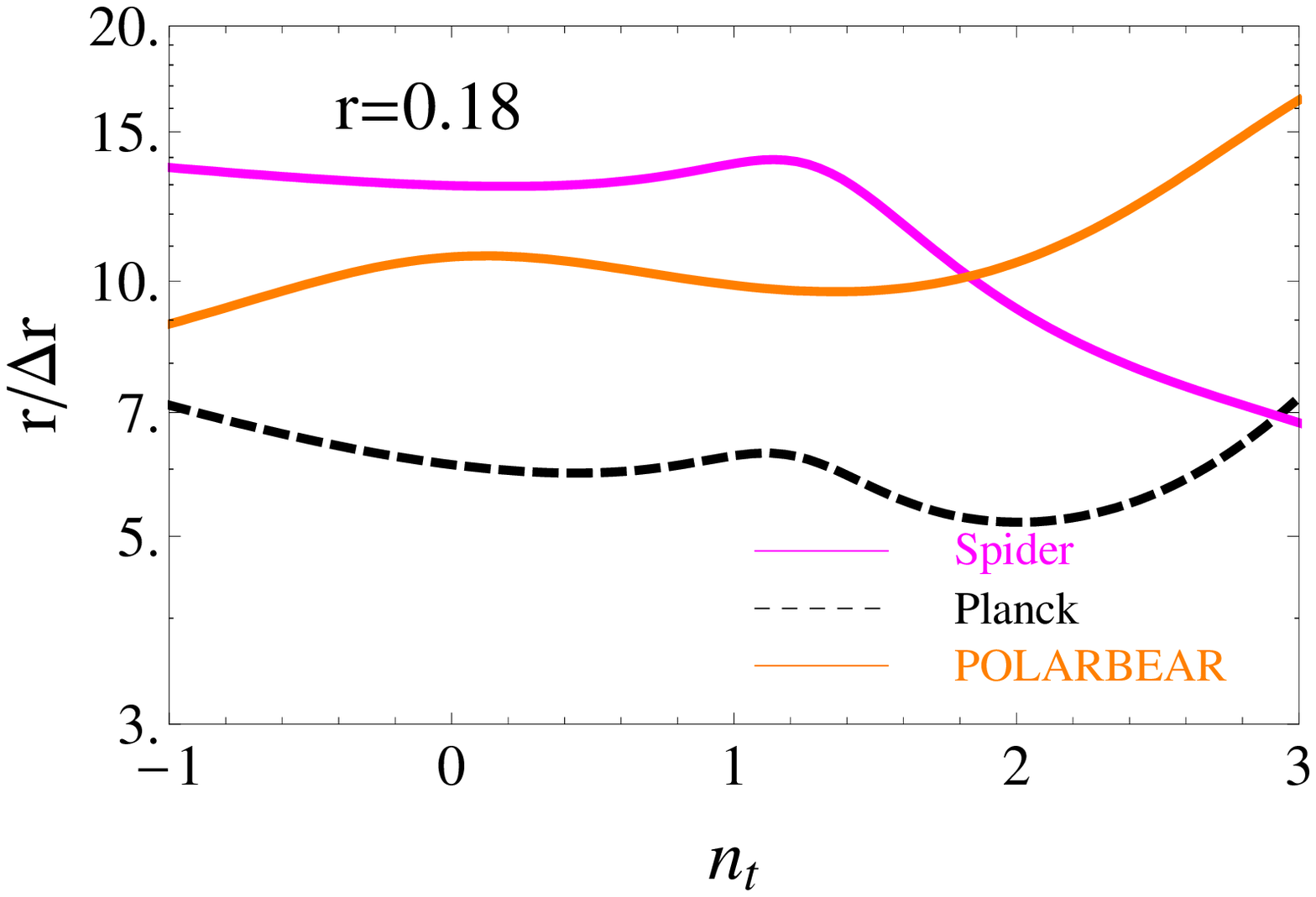}}
\caption{Predicted signal-to-noise of measurements of $r$ for
varying $r$ (left panel) and $n_{\rm t}$ (right panel). }
\label{fig:fore}
\end{figure*}

\begin{figure*}
  \centerline{
  \includegraphics[width=0.43\textwidth]{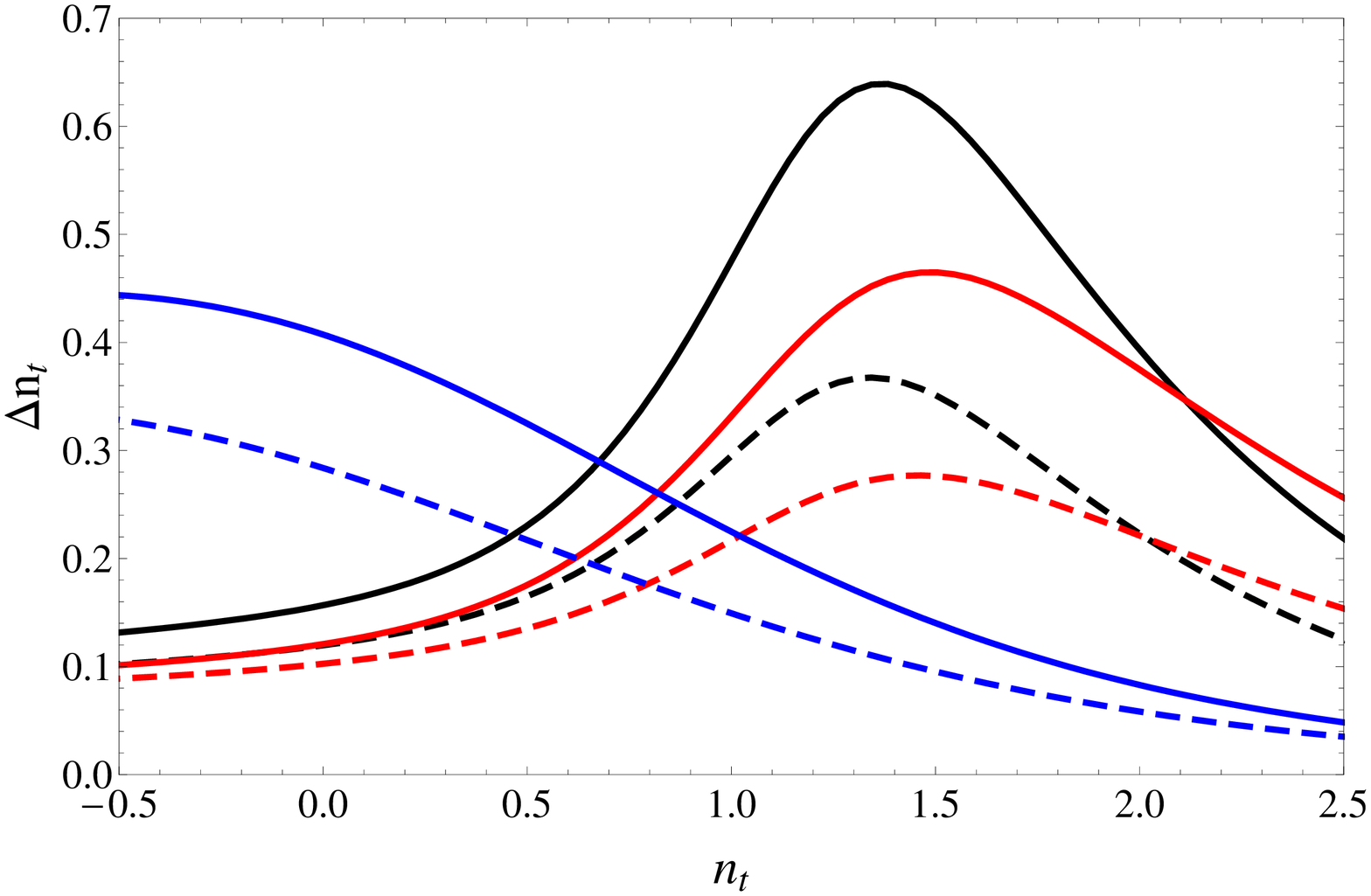}
  \includegraphics[width=0.63\textwidth]{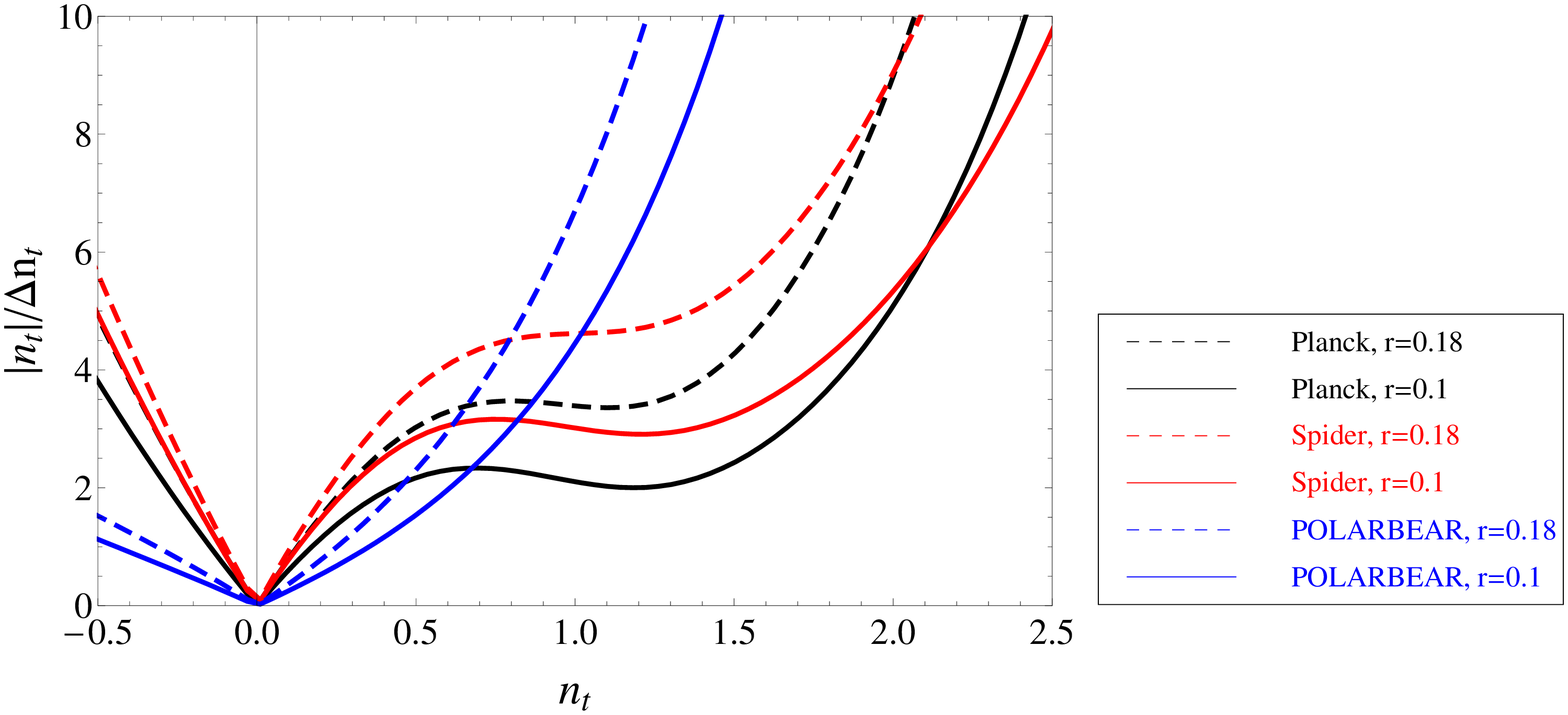}}
  \caption{Predicted $\Delta\nt$ (left panel) and signal-to-noise $|\nt|/\Delta \nt$ (right panel) for varying
  $\nt$, with $r=0.18$ and $r=0.1$.}
  \label{fig:dta-ntfig}
\end{figure*}


{\it Forecast for future experiments--} The tensor-to-scalar ratio
is claimed to be $0.18$ at $k_{0}=0.01 \,{\rm Mpc}^{-1}$ by BICEP2
experiment~\cite{BICEP2}. It now is required to determine if $\nt$
varies, how precisely can future experiments measure the B-mode
polarization. 

In Figure~\ref{fig:clbb1}, we plot the theoretical
prediction of $C_{\ell}^{\rm BB}$ with three $\nt$ values, 
using three different noise levels. The first corresponds to that of
a {\it Planck}-like experiment having the idealised noise performance described in 
Ref.~\cite{Planckblue}, and the second and third to the Spider and POLARBEAR experiments. 
We note that for the {\it Planck}-like experiment, 
we have not attempted to model the real performance of {\it Planck}, 
or the effects of systematics, and so every time '{\it Planck}' is mentioned below in the context of 
forecasted results, this refers to results from an idealised {\it Planck}-like experiment only.


We choose the three representative values of $\nt$:
(1) $\nt=0$, flat tensor spectrum (red solid line); (2)
$\nt=-r/8$, the $\nt$ value that satisfies the consistency
relation for single-field slow-roll inflation
model~\cite{Liddle93} (green solid line); (3) $\nt=2.6$, the
current $3\sigma$ upper limit of {\it Planck}+BICEP2 constraint
(blue dashed line). We can see that the $\nt=-r/8=-0.025$ line
does not differ significantly from the flat tensor spectrum.
However, as $\nt$ becomes more positive, the $C^{\rm BB}_{\ell}$
tends to have more powers on small scales and less power on large
scales, due to the blue tilted power spectrum. In addition, we
follow the recipes in ref.~\cite{Ma10} to calculate the noise
level of the each experiment. It has been seen that Spider has
lower noise than \textit{Planck} at low-$\ell$, but the effective
noise blows up at high $\ell$ because of the large beam. The noise
from POLARBAER is systematically lower than \textit{Planck} and
Spider, making it a powerful measurement on primordial tensor
mode. We also plot the gravitational lensing signal as the purple
dashed line in Figure~\ref{fig:clbb1}. The gravitational lensing
can convert primordial E mode into B mode, therefore add an
effective noise to the true primordial B-mode signal. In
Figure~\ref{fig:clbb1}, we can see that this signal peaks at $\ell
\approx 1000$ which is the typical galaxy cluster scale. The
$C^{\rm len}_{\ell}$ is fixed with a certain set of cosmological
parameters and therefore can be outputted from
\textsc{camb}~\cite{cambcode}.

In Figure~\ref{fig:clbb2}, we plot the added signal of primordial
tensor mode with gravitational lensing, and the current
measurement from BICEP2~\cite{BICEP2} and
POLARBAER~\cite{PolarBear}. We can see that current data is
consistent with the tensor mode with amplitude $r=0.18$, while it
still allows a fairly large range of spectral index $\nt$.

Assuming each $\ell$ is independent, uncertainties of each $\ell$
of B-mode polarization power spectrum is computed as:
\begin{eqnarray}
\Delta C^{\rm BB}_{\ell}=\sqrt{\frac{2}{(2\ell+1)f_{\rm
sky}}}\left(C_{\ell}^{\rm BB}+ N_{\ell}^{\rm BB} \right)
,\label{eq:delta-cl}
\end{eqnarray}
where $N_{\ell}$ is the effective noise of each experiment, which
includes the instrumental noise, residual foreground
contamination, and the gravitational lensing. The value of $f_{\rm
sky}$ is the effective area of sky that each experiment observes,
which are $0.65$, $0.5$, $0.024$ for
\textit{Planck}~\cite{Planckblue}, Spider~\cite{Crill08} and
POLARBEAR~\cite{PolarBear}, respectively.

In Figure~\ref{fig:snrcl}, we plot the signal-to-noise of each
$\ell$ for the three experiments. We can see that for each
experiment, as the $\nt$ value becomes more positive, one gains
less signal to noise from large scales, but more from small
scales. The Spider experiment, because of the large beam, cannot
obtain consistent result on large $\ell$s, but its measurement on
low-$\ell$ is better than that of \textit{Planck}. The future
POLARBEAR experiment is better than both Spider and
\textit{Planck}. We list the contribution to total signal to noise
from $\ell \leq 100$ and $\ell>100$ in Table~\ref{tab1}.

Let us forecast for the constraints achievable with future
experiments. With the assumption that each parameter is
Gaussian-distributed, we calculate the Fisher matrix
$F_{\alpha\beta}$ \cite{Tegmark97a,Tegmark97b} such that
\begin{equation}
F_{\alpha \beta }=\frac{1}{2}\textrm{Tr}[\mathbf{C}_{,\alpha
}\mathbf{C}^{-1}\mathbf{C}_{,\beta }\mathbf{C}^{-1}],
\end{equation}%
where $\mathbf{C}$ is the total covariance matrix, which includes
both signal and noise contributions. In case of B-mode only, where
each $\ell$ and $m$ mode is independent of each
other, then
\begin{equation}
\mathbf{C}_{\ell_{1}m_{1}\ell_{2}m_{2}}=(C_{\ell_{1}}^{\rm
BB}+N_{\ell_{1}}^{\rm BB})\delta _{\ell_{1}\ell_{2}}\delta
_{m_{1}m_{2}}.
\end{equation}%
In this case, the Fisher matrix can be simplified
\cite{Tegmark97a,Tegmark97b} as:
\begin{equation}
F_{\alpha \beta }=\sum^{\ell_{\rm max}}_{\ell=\ell_{\rm min}}\left( \frac{2\ell+1}{2}f_{\rm sky}\right) \frac{%
(C_{\ell}^{\rm BB})_{,\alpha}(C_{\ell}^{\rm
BB})_{,\beta}}{(C_{\ell}^{\rm BB}+N_{\ell}^{\rm BB})^{2}}.
\label{falphabeta}
\end{equation}%
For \textit{Planck} and Spider experiment, since the observation
is nearly full sky, we perform the summation in
eq.~(\ref{falphabeta}) to be $\ell_{\rm min}=2$ till $\ell_{\rm
max}=3000$. For the ground-based POLARBEAR, the summation is
performed from $\ell_{\rm min} = 21$ to $\ell_{\rm max} = 3000$,
since POLARBEAR cannot cover the largest angular scales because of
the corresponding finite survey areas.

The inverse of the Fisher matrix $F^{-1}$ can be regarded as the
best achievable covariance matrix for the parameters given the
experimental specification. The Cramer-Rao inequality suggests
that no unbiased method can measure the $i$th parameter with an
uncertainty less than $1/\sqrt{F_{ii}}$ \cite{Tegmark97a}. If the
other parameters are not known and considered as free parameters,
the minimum standard deviation is $(F^{-1})^{1/2}_{ii}$
\cite{Tegmark97a}. Therefore the best prospective signal-to-noise
ratio can be estimated as $\alpha/
\Delta \alpha$, where $\Delta \alpha =(F^{-1})_{\alpha \alpha }^{%
  {1}/{2}}$.

In the left panel of Figure~\ref{fig:fore}, we plot the $r/\Delta
r$ as a function of true value of $r$. We can see that the higher
the value of $r$ is, the more signal to noise one can obtain from
each experiment. The POLARBEAR and Spider experiments provide
stronger constraints on $r$ than \textit{Planck}. In the right
panel of Figure~\ref{fig:fore}, we vary the value of $\nt$ and
calculate the $r/\Delta r$ for each assumed value of $\nt$. It can
be seen that the for \textit{Planck}, the measured signal to noise
of $r$ is not exceedingly sensitive to the true value of $\nt$.
But for Spider, as the $\nt$ value becomes bigger, the signal to
noise decreases because Spider is incapable of measuring
high-$\ell$ power accurately. For POLARBEAR, the signal to noise
of $r$ is all high across all values of $\nt$.

In Figure~\ref{fig:dta-ntfig}, we plot the noise $\Delta\nt$ and
the signal-to-noise ratio $\nt/\Delta\nt$ respectively, for
varying $\nt$. In the left panel of Figure~\ref{fig:dta-ntfig},
$\Delta\nt$ is about $0.1$ near $\nt=0$. Thus it is still
challenging for the upcoming experiments to measure the
inflationary consistency relation (if $\nt=-r/8$). In the right
panel of Figure~\ref{fig:dta-ntfig}, It can be seen that assuming
$r\geq 0.1$ and $\nt \sim 1$, \textit{Planck} and Spider are able
to confirm the hint for positive $\nt$ and POLARBEAR will be
sufficiently precise to make a detection.

In Figure~\ref{fig:comb}, we put together the current joint
constraints with the forecasted signal-to-noise measurement of
parameters $r$-$\nt$. The green region is the excluded by current
POLARBEAR experiment at $3\sigma$ CL, while the blue and purple
contours are the BICEP2 only and \textit{Planck}+\textit{WMAP}
Polarization(WP)+BICEP2 data. It can be seen that the joint
constraint favors a positive range of $\nt$ values. We also plot
the $r/\Delta r=5$ and $3$ lines in the same figure for
\textit{Planck}, POLARBEAR and Spider experiments. Comparing
\textit{Planck} forecasted lines with the current constraints, one
can see only if $r<0.1$ and $2<\nt<3$ \textit{Planck} may not be
able to constrain $r$ at $3\sigma$ CL. In all other parameter
ranges, \textit{Planck} can constrain the value better than
$3\sigma$ CL. Specifically, if $r>0.15$, \textit{Planck} should be
able to measure it in more than $5\sigma$ CL. Spider and POLARBEAR
can do much better than \textit{Planck} since they can measure
nearly the whole parameter space with $r>0.05$ in more $5\sigma$
CL. In addition, as we can see from Figure~\ref{fig:comb}, there
is a possibility for future experiments to test the inflationary
consistency relation as shown in dashed (nearly vertical) line.
But current \textit{Planck}+BICEP2+WP data favors a large positive
value, which does not cover this line at $2\sigma$ CL. Therefore,
future experiments will set up a rigorous test on this consistency
relation.

{\it Conclusions--} The B-mode polarization power spectrum is a
unique probe of the primordial tensor fluctuations. Current
observations BICEP2 claim that the tensor-to-scalar ratio $r$ is
$0.18$ at $0.01\,{\rm Mpc}^{-1}$. If BICEP2 data is correct, $\nt$
is constrained to be $\nt=1.24\pm0.90$ ($1\sigma\,$CL) for BICEP2
only, indicating a blue tensor spectrum. By combining the BICEP2
data with \textit{Planck} data and \textit{WMAP} polarization
data, we find $\nt = 1.76 \pm 0.54$ ($1\sigma\,$CL).

Assuming the true value of $r$ is large and detectable, we
forecast the detectability of the parameters $r$ and $\nt$ for a
\textit{Planck}-like experiment with the same noise as
projected in Ref.~\citep{Planckblue}, and for
balloon-borne Spider data and ground-base POLARBEAR data with the
current understanding the foreground emission and their
experimental noise. We used the Fisher matrix to calculate the
forecasted signal-to-noise ratio. We found that if $r>0.1$ and
$2<\nt<3$, \textit{Planck} can measure $r$ in more than $3\sigma$
confidence level. POLARBEAR and Spider data are even more powerful
than \textit{Planck}, since they can measure nearly the whole
parameter space $r>0.05$ by more than $5\sigma$ CL. The
detectability of tensor-to-scalar ratio $r$ for \textit{Planck},
Spider and POLARBEAR is relatively independent on the details
value of $\nt$ since the $r/\Delta r=3$ and $5$ lines in
Figure~\ref{fig:comb} are nearly horizontal. However, we caution
the reader that if the BICEP2 result is found to be largely due to
uncleaned polarized foreground, the true value of $r$ could be
significant less than $0.2$. In this case, the signal-to-noise
lines of $r$ in Figure.~\ref{fig:comb} will be very low and might
be undetectable. In addition, in our study, we do not consider the
running of spectral index ($\der \nt/\der \ln k$) for tensor power
spectrum, which in principle, can be nonzero if the spectral index
is large. In addition, successful delensing can significantly
boost the signal-to-noise ratio, particularly when $\nt$ is
positive~\cite{Seljak04}, which is beyond the scope of this paper.

\bigskip
{\it Acknowledgments--}
We thank BAUMANN D.~ and ZHAO W.~ for helpful discussions. This work was supported by a CITA National
Fellowship, a Starting Grant of the European Research Council (ERC STG Grant No. 279617) and the Stephen Hawking Advanced Fellowship.
Part of this work was undertaken on the COSMOS Shared Memory system at DAMTP, University of Cambridge operated on behalf of the STFC DiRAC HPC Facility. This equipment was funded by BIS National E-infrastructure capital grant ST/J005673/1 and STFC grants ST/H008586/1, ST/K00333X/1.


\end{document}